# Magnetosphere Magnetic Field Wobble Effects on the Dynamics of the Jovian Magnetosphere

R. M. Winglee[1] and E. M. Harnett[1]

(1) Department of Earth and Space Sciences

University of Washington

Seattle, WA 98195-1310

Correspondence to: R. M. Winglee (winglee@uw.edu)

**Abstract.** The Jovian magnetosphere is complicated by the multiple plasma sources and ion species present within it. In addition, it is a fast rotating system with its dipole axis titled from its rotational axis. To date global models of Jovian have neglected the presence of the different ion species as well as the tilted nature of the dipole axis. This paper reports the results of the first multi-fluid global modeling of these effects in a single self-consistent study for processes occurring in the outer magnetosphere. In the inner magnetosphere the model densities are shown to be comparable to observed densities with much of the density variables due to the wobble. The wobble enables plasma to be transported to higher latitudes and then centrifugal acceleration leads to radial transport of the plasma. Even out to 100 $R_J$ the dominant source of plasma is from the Io torus and not the solar wind. Embedded in the signatures of the plasma properties is the wobble of the magnetic field. At the interface between the middle and outer magnetosphere, the wobble produces a sinusoidal modulation of the plasma properties, which yields at a fixed observing point two density peaks, each with the planetary period. However, because of solar wind forcing and the fast rotation of Jupiter, torques are exerted on the outer magnetospheric plasma such that the azimuthal speed between the southern and northern hemispheres are unequal. This leads to the asymmetric square wave modulation of the plasma properties in the outer magnetosphere such that only one density peak is seen each planetary period in the outer magnetosphere. In addition, this differential speed leads to reconnection in the tail at the planetary period and enhanced ejection of plasma down the tail at the planetary period. These processes could explain the periodicity seen in the New Horizons' data without the need for invoking arbitrary solar wind conditions. It is also shown that the periodicity is not sensitive to the solar wind density but the particle energization and the amplitude of the flapping of the tail current sheet is dependent on the solar wind conditions. This dependence on the solar wind conditions thereby modifies the probability of observing the reconnection events if the spacecraft is displaced from the rotational equator. The energetic particles generated by these processes are shown to map to the arc-like structures seen in the dusk side auroral features.

1. **Introduction**

The Jovian magnetosphere is probably the most complex planetary magnetosphere within the solar system. This complexity arises due to the presence of the multiple sources of plasma and different ion species within the Jovian magnetosphere and a host of processes that can energize the plasma in different regions of the magnetosphere. The inner magnetosphere (R < 10 $R_J$ where 1 $R_J$ =71,492 km is the radius of Jupiter) is strongly controlled by the fast rotation of Jupiter's magnetic field, and the plasma is approximately co-rotational. The plasma is predominantly supplied by the ionization of $SO_2$ that originates from Io at an overall rate of 1 ton/s or about $6 \times 10^{29}$ amu/s (or about $10^{28}$ ions/s) (*Bagenal and Sullivan*, 1981; *Hill et al.*, 1983; *Vasyliunas*, 1994). This ionization introduces several heavy ions species including $O^+$, $S^+$, $O^{++}$, $S^{++}$ and $SO_2^+$ or $S_2^+$ within the Io plasma torus (*Bagenal*, 1994), which has a peak density in excess of 1000 $cm^{-3}$. The magnetic field is sufficiently strong that the plasma in this region is approximately in co-rotation with the planet.

The Io plasma torus itself is highly structured with three distinct regions: a) The inner "cold" torus lies between about 5 and 5.5 $R_J$ and a height of < 0.1 $R_J$ off the equator and temperature of a few eV, b) the ribbon which lies between about 5.6 and 5.8 $R_J$ with a height of ~0.6 $R_J$ and where the $S^+$ density spikes to densities above 1000 $cm^{-3}$, and c) the warm plasma torus which lies beyond the ribbon and with a temperature of 20-120 eV (*Herbert et al.*, 2008). The supply of plasma to these regions appears to be in part produced by the local pickup of iogenic ions in close proximity to Io at about 200 kg/s and the remaining 800 kg/s from an extended region of ionization distant from Io (*Wilson et al.*, 2002). Multi-species interactions at Io and in the torus are important in controlling the total number density and mass density (*Delamere et al.*, 2004, *Dols et al.*, 2008).

The interaction of Io with the torus plasma is sufficiently strong that auroral emissions are observed at Jupiter in the vicinity of the magnetic footpoint of Io (*Connerney et al.*, 1993). The footpoint auroras have been shown to be highly structured, with distinct leading and following intensifications, or spots. The relative positions and intensities of these spots are seen to vary in time and are correlated with the position of Io relative to the center of the plasma torus (*Gérard et al.*, 2006; *Serio and Clarke*, 2008) with the largest interval between spots occurring in the hemisphere were Io is closest to the edge of the torus. An additional auroral spot is located alternatively upstream or downstream of the main Alfvén wing spot (*Bonfond et al.*, 2008) with the lead spot tending to appear in the opposite hemisphere where the spots are most prominent. The lead spot can be 1-10 Io diameters in length while the tail can be as long as 100º in longitude (*Clarke et al.*, 2002; *Gérard et al.*, 2006). Europa's footpoint emissions are much weaker and can only be detected out to about 70 Europa diameters (*Clarke et al.*, 2002; *Gérard et al.*, 2006). Ganymede's is much brighter than Europa's but probably only extends out to 20 Ganymede diameters (*Clarke et al.*, 2002; *Grodent et al.*, 2009).

The next region, the middle magnetosphere, is between 10-40 $R_J$. The magnetic field is insufficient to enforce co-rotation for the plasma in this region (*McNutt et al.*,1979). The position of the loss of co-rotation occurs at about 18 $R_J$ and this position is consistent with a mass loading of about $10^{30}$ amu/s (*Hill et al.*, 1980). The breakdown of co-rotation has been proposed as one of the main drivers of Jupiter's auroral emissions, which map into the middle magnetosphere (*Hill*, 2001; *Saur et al.*, 2003).

The middle magnetosphere is also a very dynamic region. In addition to iogenic $O^+$ and $S^+$ being present in the region, there are significant components of $H^+$ and $He^+$ which suggest that at least some of the plasma is of solar wind origin in the regions. The difference in the ratio of H+/He+ from the solar wind value though suggests that there is also a significant non-solar origin of the protons as well (*Mall et al.*, 1993). There are also local sources of plasma from the icy Galilean moons with Europa and Ganymede providing about $6\times10^{26}$ ions/s and Callisto about $5\times10^{26}$ ions/s (*Kivelson et al.*, 2004) with the interactions at Europa and Ganymede leading to the discernible aurora at their magnetic footpoints in Jupiter's topside atmosphere, already discussed (*Clarke et al.*, 1998). Much of the plasma in this region is confined within a thin current sheet with a width of 2-3 $R_J$ (*McNutt et al.*, 1981).

The outer magnetosphere at > 40 $R_J$ can also be highly active. For example, distinct plasma structures moving down the tail have been observed at distances of over 1000 $R_J$ (*McComas et al.*, 2007; *McNutt et al.*, 2007). These plasmoids have a mixture of Io and Jovian ionospheric plasma with 10 hr periodicity, along with intense injections of Jovian ionospheric plasma with a period of about 3 days. These events have been proposed to be related to reconnection events (*McNutt et al.*, 2007; *Krupp*, 2007) possibly driven by reconnection processes in the outer magnetosphere. However, the exact source mechanism for these ejection events has yet to be determined.

These different regions are not independent but are coupled by a variety of processes. One of these processes is the interchange instability where mass heavy flux tubes in the inner magnetosphere are exchanged with hotter more tenuous plasma from further out. Evidence for small scale interchange processes have been reported by *Bolton et al.*, (2007), *Kivelson et al.* (1997) and *Thorne et al.* (1997) while large scale injections of hot tenuous plasma from the middle magnetosphere have been reported by *Mauk et al.*, (1999). These injections occur at all System III longitudes and local time positions and add further to the mixing and energization of different plasmas and ion species but have only been observed in the inner magnetosphere. *Kronberg et al.* (2008) suggest that periodic plasmoid ejections are responsible for the observed energetic particle injections. It is worth noting that interchange instabilities are very much more evident in the Kronian magnetosphere (*Mauk et al.*, 2005) and appears within multi-fluid global modeling of the Kronian magnetosphere (*Kidder et al.*, 2009).

The complexity of the Jovian magnetosphere has made the development of a comprehensive quantitative model very difficult. There have been a number of global MHD simulations of the Jovian magnetosphere (*Ogino et al.*, 1998; *Miyoshi and Kusano*, 2001; *Walker et al.*, 2001; *Walker and Ogino*, 2003; *Fukazawa et al.*, 2005, 2006; *Moriguchi et al.*, 2008). These global simulations have focused on the influence of solar wind conditions on the dynamics within the Jovian magnetosphere, and have quantified the changes in the outer magnetosphere under the influence of parallel and antiparallel interplanetary magnetic field (IMF). However, the inner radius of these simulations is typically greater than 15 $R_J$ except for the most recent simulations by *Moriguchi et al.*, (2008) where the inner radius was moved to 8 $R_J$. Even so, none of these simulations actually includes the torus, which is the main contributor of the plasma to at least the inner and middle magnetosphere. *Chané et al.* (2013) addressed this issue by including torus and ionospheric coupling through a physics module, though inner boundary remained at 10 $R_J$. It should also be noted that these models neglect the wobble of the Jovian magnetic field. To generate the periodicity seen in energetic particles down the tail observed by *McComas et al.*, (2007) and *McNutt et al.*, (2007), interplanetary magnetic field (IMF) rotation at the planetary period were required in the simulations by *Fukazawa et al.*, (2010). The presence of this rotation of the IMF has yet to be corroborated by observations, particularly over the very large period of observations from New Horizons.

In this paper we report results from a multi-fluid global model of the Jovian magnetosphere (Section 2) that has the inner radius at 2 $R_J$, enabling the incorporation of the Io plasma torus and the wobble into the simulation model. The derived properties of the inner magnetosphere are discussed in Section 3. It is shown that the overall properties are in approximate agreement with observation, though the model shows that due to the wobble enhanced transport to high latitudes occurs and when this plasma experiences centrifugal acceleration it leads to super-corotational flows in the equator near the breakdown of co-rotation. We demonstrate here that the wobble, along with the fast rotation of Jupiter, leads to the development of quasi-periodic ejection of plasmoids at the planetary period. Sections 4 and 5 discuss the influence of the magnetic field wobble (previously neglected in other models) on the production of large scale flux ropes, and how the mixture of Io torus plasma relative to solar wind and ionospheric ions is modulated on a time scale related to the planetary rotational period. It is shown that for this reason, it is important to include the different ion species and different sources within the Jovian magnetosphere, as well as the wobble of the magnetic field. While the periodicity is not sensitive to changing solar wind conditions the development of energetic populations in the magnetosphere is demonstrated to have observable effects in both the distant tail (Section 5) and in auroral process (Section 6). A model for the importance of including magnetic wobble is given in Section 7 where torques on the plasma from the wobble and fast rotation are proposed as the driving influence for the above processes.

## 2. Simulation Model

The results shown in this paper employ a 3D global model for the Jovian magnetosphere that includes: the interaction of this large magnetosphere with the induced magnetospheres of Io and Europa and the small magnetosphere of Ganymede; and the influence of different ion species such as $S^+$, $O^+$, $S^{++}$ and $H^+$ from the different plasma sources within the magnetosphere.

The model has substantial heritage that includes the first coupled model for Saturn/Titan interactions that includes multiple ion species with variable gridding within the magnetosphere and moving higher resolution gridding about the moon to incorporate its rotation about the planet (*Winglee et al.*, 2013). Elements of the model have also been used to examine weakly magnetized systems including Mars, (*Harnett and Winglee*, 2003, 2007), Mercury, (*Kidder et al.*, 2009), Ganymede *(Paty et al.*, 2004, 2006), and Titan (*Snowden et al.*, 2007, 2013) The multi-fluid model has been used to planetary magnetospheres of the Earth (*Winglee*, 2004; *Winglee et al.*, 2002, 2005) and the Kronian magnetosphere (*Kidder et al.*, 2009, 2012).

The equations that are used in the modeling are from *Chapman and Cowling* (1952), and as re-derived by *Siscoe* (1983). Specifically, the evolution of the distribution function, *f,* for all collisionless plasmas is governed by the Vlasov equation:

$$\frac{\partial f}{\partial t} + \mathbf{v} \cdot \frac{\partial f}{\partial \mathbf{x}} + \mathbf{a} \cdot \frac{\partial f}{\partial \mathbf{v}} = 0 \qquad (1)$$

$$\mathbf{a} = \frac{q}{m}[\mathbf{E} + \mathbf{v} \times \mathbf{B}] \qquad (2)$$

where $v$ and $a$ are the particle velocity and the acceleration of the particle, respectively. By definition the density $n$, and pressure tensor $P_{ij}$ are given by

$$\int f d^3 v = n(x,t) \tag{3}$$

$$\int v f d^3 v = V(x,t) n(x,t) \tag{4}$$

$$\int w_i w_j f d^3 v = P_{ij} \tag{5}$$

where $v$ is the velocity of each individual particle and $w = V-v$ is the particle thermal velocity.

The multi-fluid approach, like hybrid models, treats the electrons as a fluid. Assuming the electrons are in drift motion (i.e. no acceleration) and that the electrons have an isotropic temperature, the electron properties are described by the modified Ohm's law and pressure equation

$$E + V_e \times B + \frac{\nabla P_e}{e n_e} = 0 \tag{6}$$

$$\frac{\partial P_e}{\partial t} = -\gamma \nabla \cdot (P_e V_e) + (\gamma - 1) V_e \cdot \nabla P_e \tag{7}$$

where $\gamma$ is the polytropic index. The magnetic field is given by

$$\frac{\partial B}{\partial t} + \nabla \times E = 0 \tag{8}$$

Assuming quasi-neutrality, one can derive the remaining equations

$$n_e = \sum_i \frac{q_i n_i}{e}, \quad V_e = \sum_i \frac{q_i n_i}{e n_e} V_i - \frac{J}{e n_e}, \quad J = \frac{1}{\mu_0} \nabla \times B \tag{9}$$

Substitution of (9) into (6) yields the standard form of the modified Ohm's law:

$$E = -\sum_i \frac{q_i n_i}{e n_e} V_i \times B + \frac{J \times B}{e n_e} - \frac{1}{e n_e} \nabla P_e \tag{10}$$

In the multi-fluid codes, the bulk moments are first derived using (3) and (4) and then simplifying approximations on (5) the various pressure equations. Taking the zeroth (i.e. $\int d^3 v$) and first moments (i.e., $\int d^3 v\, v$) of (3) yields the equations for each ion species α, for the mass density $\rho_\alpha$ and bulk velocity $V_\alpha$

$$\frac{\partial \rho_\alpha}{\partial t} + \nabla \cdot (\rho_\alpha V_\alpha) = 0 \tag{11}$$

$$\rho_\alpha \frac{dV_\alpha}{dt} = q_\alpha n_\alpha (E + V_\alpha \times B) - \nabla \cdot \mathbf{P}_\alpha \tag{12}$$

where $\mathbf{P}_\alpha$ is the full pressure tensor. The pressure tensor is related to the temperature tensor such that $\mathbf{P}_\alpha = n\, k\, \mathbf{P}_\alpha$ where k is Boltzmann's constant. The 2$^{nd}$ moment yields the pressure tensor equation given by (*Chapman and Cowling,* 1952; *Siscoe,* 1983):

$$\frac{\partial \mathbf{P}_\alpha}{\partial t} = -\mathbf{V}_\alpha \cdot \nabla \mathbf{P}_\alpha - \mathbf{P}_\alpha \nabla \cdot \mathbf{V}_\alpha - \mathbf{P}_\alpha \cdot \nabla \mathbf{V}_\alpha - (\mathbf{P}_\alpha \cdot \nabla \mathbf{V}_\alpha)^T + \frac{e}{m_\alpha}[\mathbf{P}_\alpha \times \mathbf{B} + (\mathbf{P}_\alpha \times \mathbf{B})^T] - \nabla \cdot Q \quad (13)$$

where Q is the heat flux term. For this treatment, the heat flux term is neglected, similar to the double adiabatic treatments. This is a reasonable assumption if the system is strongly controlled by convection, as in the global magnetosphere. All the other elements of the pressure tensor are retained which means that the total number of equations for each ion species consists of one density equation, three velocity equations and 6 pressure equations. This set of equations requires substantially more computer resources than the total of 5 equations for the plasma dynamics of MHD and the total of 6 equations used in the double adiabatic treatment. The isotropic pressure equations can be derived from the above equations by assuming $P_{ij} = 0$ if $i \neq j$ and $P_{ij} = P$ for $i = j$, while the double equations can be derived by assuming

$$P_{ij} = P_\| \hat{b}_i \hat{b}_j + P_\perp (\delta_{ij} - \hat{b}_i \hat{b}_k) \ . \quad (14)$$

The main effect that is seen with the inclusion of anisotropy is that the loss rate of plasma along the field lines into the Jovian magnetosphere is reduced by about half. These are the focus of a separate paper.

In order to cover the full Jovian magnetosphere, we use a stacked set of grids where the grid resolution varies by a factor of 2 between successive grids. All grids are run simultaneously, and information is passed between grids at each time step, i.e. the plasma and fields from the finer grid is interpolated onto the coarser grid while the coarser grid provides the outer boundary conditions for the finer grid. The nested grids have a resolution of 0.2 $R_J$ around Jupiter and out to 12 $R_J$ magnetosphere. The inner boundary is set at 2.0 $R_J$. The resolution then increases by a factor of 2 between successive grids at larger radial distances from Jupiter. With this scheme, the model is able to cover more than a 1000 $R_J$ down the tail and several hundred $R_J$ across the tail. At the outer regions the resolution is about 12 $R_J$. The coordinates in the figures are such that X points away from the Sun, Z is points along the magnetic north pole, and Y completes

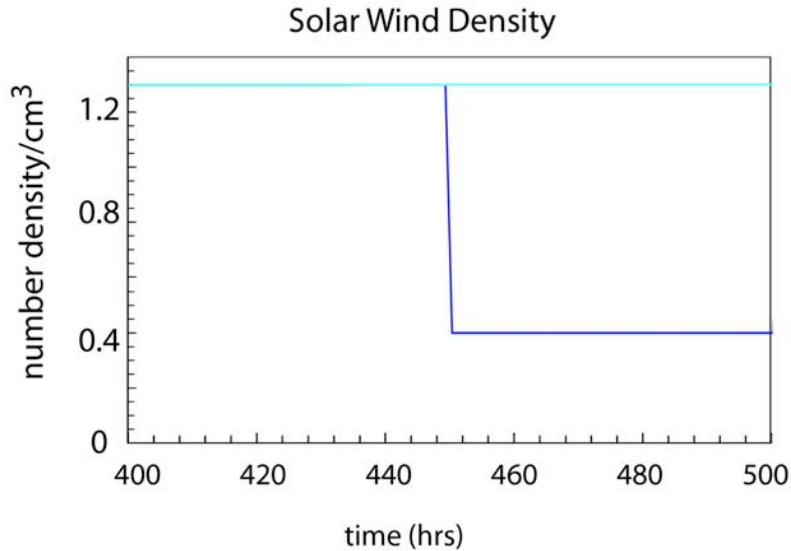

**Figure 1.** Solar wind density profile for simulations presented in this paper. A high (light blue) and a low density (dark blue) conditions were assumed to test the sensitivity of the results to the prevailing solar wind conditions.

the system.

The model requires a significant ionospheric density in order to keep the Alfvén speed at the inner boundary sufficiently small allow reasonably large times steps but not so small as to modify the overall densities in the magnetosphere. The ionospheric density is set empirically by performing test runs and determining the proton density in the inner and middle magnetosphere relative the density reported in *Bagenal* (1994). Our tests indicate that an overall ionospheric density of 400 cm$^{-3}$ over the polar regions, at the inner boundary, is sufficient to produce the observed H$^+$ densities in the magnetosphere. This density yields an Alfvén speed of $10^5$ km/s over the poles so that there are no speeds in the simulations that exceed the speed of light. The density is assumed to them fall off by a factor of 2 towards the equator, to enable a differentiation of polar and torus plasma sources. Our assumed density is consistent with the density profile of *Yelle and Miller* (2004) assuming a $1/R^3$ falloff from the ionosphere. The density though is much higher than inferred from radio emissions, e.g. *Zarka et al.*, (2001), but these are associated with the auroral acceleration region, where the number density drops as the plasma tries to conserve total particle flux. The same is true in the terrestrial auroral region, where the overall plasma density is higher than that inferred from the auroral acceleration region.

Our testing of the simulation model indicates that the results are not strongly dependent on the ionospheric resistance, with the dominant driver the Io torus parameterization. In the following we set the ionospheric resistance similar to the terrestrial ionosphere. Specifically, the region within the inner boundary is given a finite resistance equivalent to a Reynolds number of 1000. At the actual inner boundary (representing the ionosphere), the Reynolds number is increased to 2000 and at one grid point above it is set at 4000. At all other points the resistivity is zero. These values yield an overall height integrated resistivity of about $10^5$ Ohm m$^2$. Because of the large area involved the total resistance is small at about 1 μOhm.

In order to sustain the observed Io plasma torus properties, we have a continuous injection of O$^+$ & S$^{++}$ at $8 \times 10^{28}$ ions/s (or 2000 kg/s), S$^+$ at $1 \times 10^{28}$ S$^+$ ions/s (or 500 kg/s) and H$^+$ at $4 \times 10^{28}$ (or 60 kg/s) to yield densities close to those of *Kivelson et al.* (2004). This injection rate is about 2-3 times higher the inferred injection rates at Io. The reason for the higher injection rate is that because the model has its inner boundary at 2 R$_J$, which leads to nearly half the injected plasma being lost to the inner boundary. As a result, the outward radial diffusion rate is about half that of the injection rate and this radial diffusion rate is comparable to the observed diffusion rate. A more detailed discussion of the loss rate is given in Section 3.

The code was run to equilibrium after 400 hrs real time. During this period there was some variation in the solar wind conditions as well as the torus loading as mentioned above. We found that for realistic variations of the interplanetary magnetic field (IMF), the magnetosphere showed very little sensitivity to the IMF, and that the dominant effect comes from the solar wind dynamic pressure. In the following, we present results from the simulations beginning at t = 440 hrs and going to t = 500 hrs. The solar wind speed was kept constant at 450 km/s during this period as well as for the preceding 100 hrs. In one series, the solar wind density was kept constant at 1.4 particles/cm$^3$ starting at t = 400 hrs. In order to demonstrate sensitivity of the results to solar wind dynamic pressure, the solar wind density was dropped to 0.4 particles/cm$^3$ shortly after T = 450 hrs, as shown in Figure 1. The IMF included a B$_y$ component at 0.1 nT and B$_z$ was varied on periods exceeding the planetary period between +/- 0.05 nT with an average of 0 nT.

## 3. Properties of the Inner Magnetosphere.

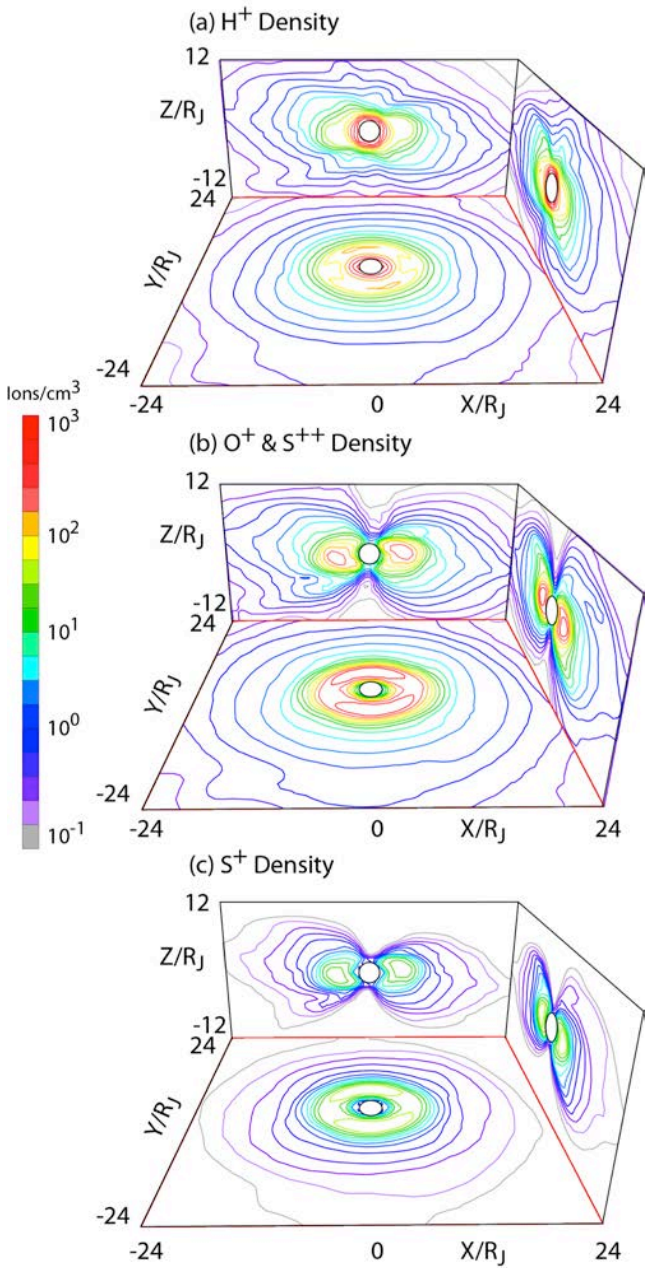

**Figure 2**. The derived density profiles for the three species ($H^+$, $S^{++}$ & $O^+$ and $S^{++}$) incorporated in the model. The black circles indicate the inner radius. Contours are shown the equatorial plane (center panel), noon-midnight meridian (back panel) and dawn/dusk terminator (right panel). A dense plasma torus is evident out to several $R_J$. A tenuous torus with high latitude extension is present to well passed Calisto's orbit.

Figure 2 shows a snapshot of the density profiles from the simulation model. The density profiles in the inner magnetosphere are not sensitive to solar wind conditions so only the data for the high solar wind case is shown. A high density torus with densities greater than 10 cm$^{-3}$ is seen out to about 15 $R_J$ and extending only about 3 $R_J$ above the equatorial plane. Beyond this region is a more tenuous torus, which connects to the higher latitude polar regions. The high latitude extent of this region indicates that the transport is not purely radial and that transport can occur by the cold inner torus plasma being transported to higher latitudes first and then centrifugally accelerated out into the magnetosphere. Since the distance to travel across the field lines is smaller than pure radial diffusion, this process represents a significant loss of plasma form the inner torus in the present model.

To more quantitatively examine density characteristics, Figure 3 shows the time profiles of the densities at positions along Io's, Europa's and Ganymede's orbits. In the vicinity of Io's orbit the peak density in the model is about 1200 as one moves through the center of the torus. This peak is about half the maximum value reported by *Kivelson et al.* (2004) and this difference is due to the fact that the model does not include a discrete Io source. The minimum value is consistent with *Kivelson et al.* (2004). At Europa, the total density lies between 30 and 70 cm$^{-3}$ with the peak value about half of that given by *Kivelson et al.* (2004). At Ganymede, the model density is between 1 and 8 cm$^{-3}$, which agrees well with *Kivelson et al.* (2004). Note that the bulk of the density variation is driven by the wobble of the torus through the observing position. The relative agreement between the model and the observations indicates that the radial and latitudinal extent of the model is in reasonable agreement to within factors of 2 to the observations.

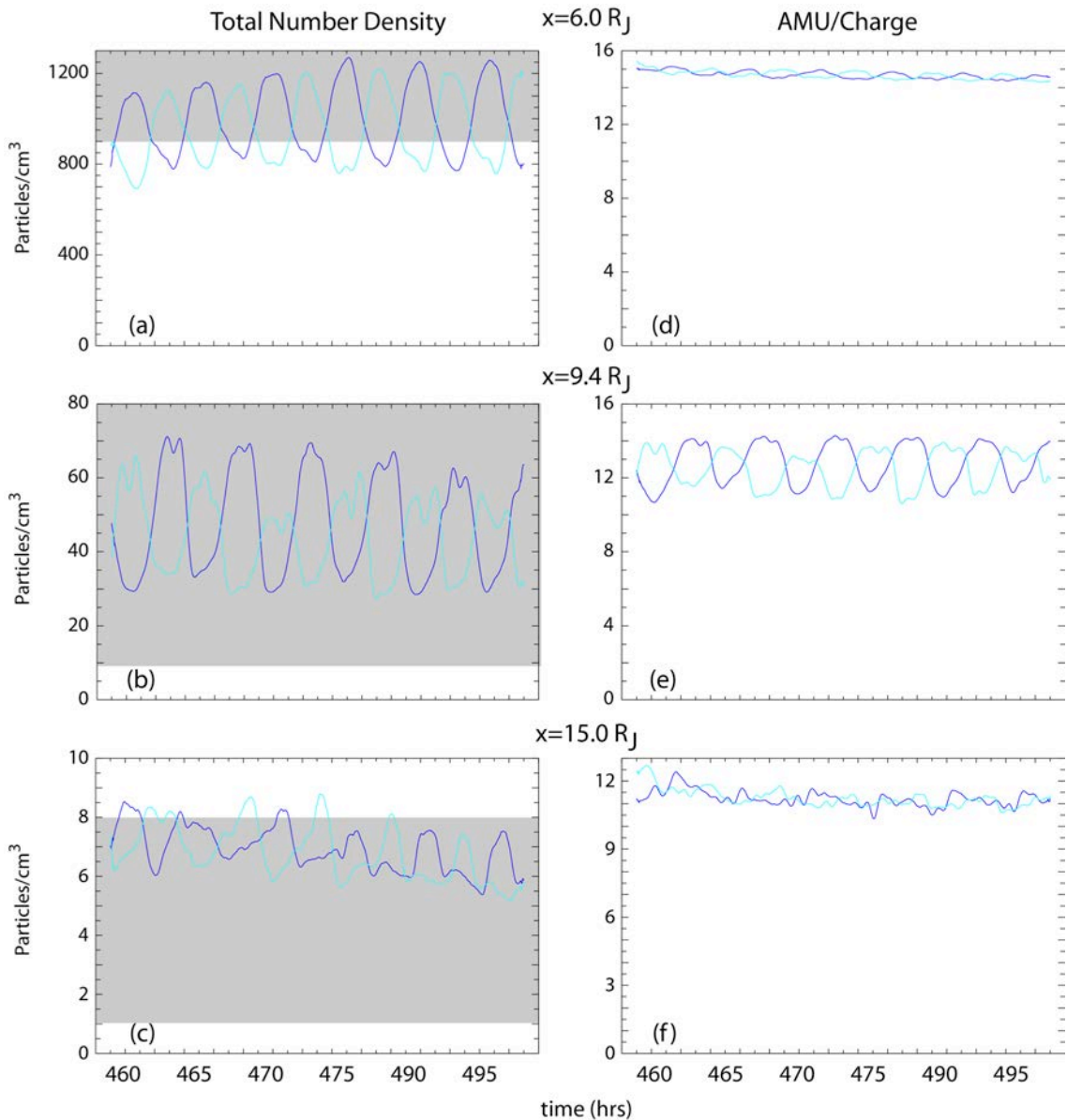

**Figure 3**. Derived density profiles (left) at the radial distance of Io, Europa and Calisto and atomic mass unit per unit charge (AMU/Charge, right). The dark blue line is for midnight local time and light blue at dawn at Io and Calisto and dusk for Europa. The gray rectangles indicate the density range as compiled by *Kivelson et al.* (2004).

In addition to calculating the total density, we can determine whether the composition of the plasma is correct by determining the AMU of the plasma at the different regions. However, since the simulation model cannot differential between $S^{++}$ and $O^+$ in terms of plasma dynamics, the right hand side of Figure 3 gives the AMU per charge. Around Io the model has a AMU/charge of 15. The reported average ion charge of 1.3 would yield an average AMU of 19.5, which is within the range as given by *Kivelson et al*. (2004). At Europa, the model AMU/charge has a range of 14 near the center of the torus and a value of 11 near the lobe. Again using an average ion charge of 1.3 would yield an AMU of 18.2 to 14.3. The peak value is consistent with *Kivelson et al*. (2004) but the lobe value is smaller than the 17 AMU reported by *Kivelson et al*. (2004). At Ganymede's orbit, there is significantly less variation in the AMU/charge, with a value of about 11.5 +/-0.5. The corresponding AMU of 15 is close to the reported value of 14 reported by *Kivelson et al.* (2004). Thus the overall variation in composition predicted by the model appears to be in reasonable agreement to with the observations.

In order to examine the energization of this plasma, Figure 4 shows the bulk velocity normalized to the co-rotation velocity and the ion sound speed Mach number and the Alfvénic Mach number corresponding to the density profiles in Figure 2. The plasma starts as co-rotational at the inner boundary and then experiences a slow down about Io as the torus plasma is added to the system. The plasma is then slowly spun up but the final speed is dependent on local time. The highest speeds appear near midnight and dawn due to the addition of the rotation and a magnetosphereic two-cell convection pattern. Mid-afternoon to dusk sector has speeds that remain below or near co-rotation speed.

These velocities are higher than observations where the bulk velocity tends to lie below the co-rotation velocity. In the model, they originate because the plasma is not confined to the equator due to the presence of the wobble. High latitude plasma experiences centrifugal acceleration as the plasma moves towards the equator, which produces the extra velocity that we see in the simulations. This additional acceleration is seen in the ion sound Mach number where the cross-cuts show additional heating along the leading edge of the high latitude field lines of the torus. The model does reproduce the observations that the plasma near Io is supersonic and near Mach 1 at Europa. The model does have the plasma still near Mach 1 near Ganymede whereas the observations have the plasma sound speed at Mach 0.8. The plasma remains below the Alfvén speed in the radial direction until after the peak in the azimuthal velocity is attained. The transition from sub-Alfvénic to super-Alfvénic flows around Ganymede's orbit is consistent with the observations of *Kivelson et al.* (2004).

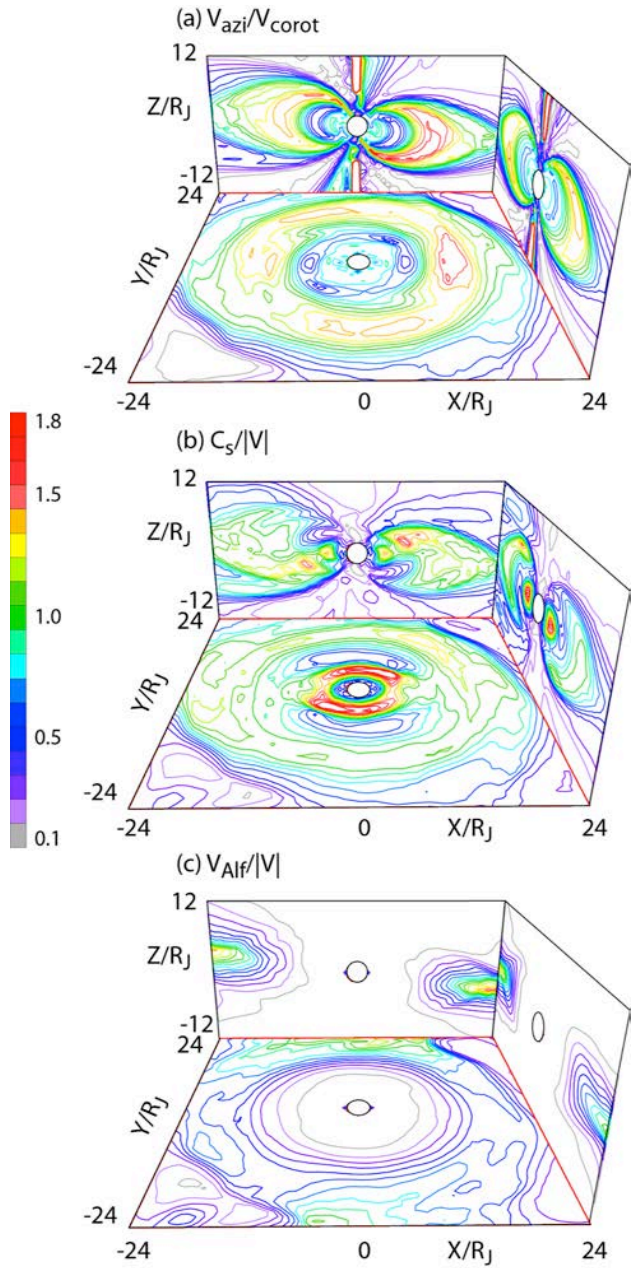

**Figure 4**. (a) The azimuthal velocity relative to the co-rotational velocity, (b) the ion sound Mach number and (c) the Alfvénic Mach number, with the same format as Figure 2.

The loss rate of plasma down the tail is shown Figure 5. This tail loss rate is about half that of the injection rate, with the other half of the plasma presumably being lost to the inner boundary. The reason for this loss mechanism is seen in the density plots of Figure 2 where the high latitude extensions of the plasma torus reaches the inner boundary of the simulations which then imposes a reduced heavy ion density. The overall net loss rate down the tail of approximately $4 \times 10^{28}$ heavy ions/s, yields the expected loss rate of about 1000 kg/s. Note though that the $H^+$ loss rate is approximately equal to the $S^{++}$ & $O^+$ loss rate despite the fact that it is injected at half the rate of the $S^{++}$ & $O^+$. The model does not separately track the origin of the $H^+$ ions (i.e. torus or ionospheric) but the ratio of the loss rate to injection rate suggests that about half the $H^+$ ions are presumably coming from the assumed ionospheric conditions.

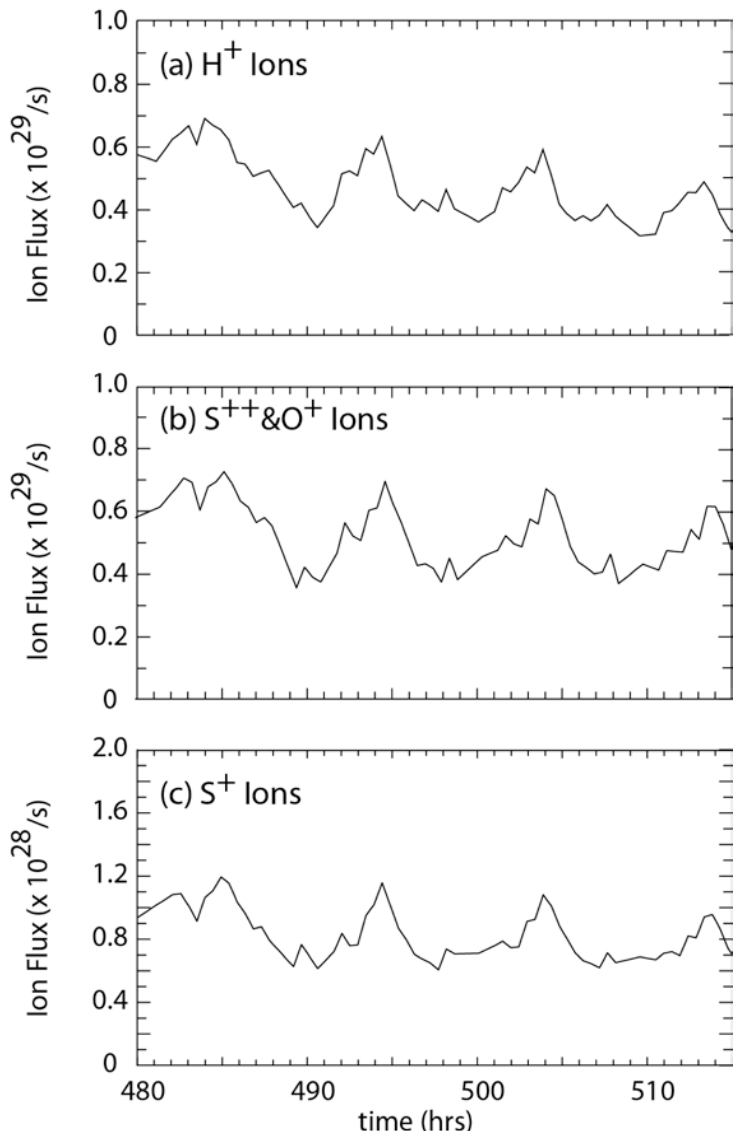

**Figure 5.** Time history of the plasma flux moving down the tail. The particle flux is seen to have a periodicity near the plasma period.

The other important feature is that the loss rate has periodic enhancements near the planetary period. This periodicity means that there is an asymmetry between the northern and southern hemisphere as were they the same, the loss rate would have a period of half the planetary rotation period. In the next sections we investigate the properties of the plasma flux moving down the tail.

**4. Plasma Periodicity Signatures in the Outer Magnetosphere.**

Figure 6 shows the density and AMU/charge histories for three different locations moving down the tail. At Calisto's position (top two panels), the total number density varies from about 0.4 to 1.6, with the higher solar wind dynamic pressure case having the higher total number density. This total number density comprises approximately 50% $O^+/S^{++}$, 38% $H^+$ and 12% $S^+$ ions to yield an AMU/charge of between 9 and 12. The higher (lower) values of the AMU are seen in association with the higher (lower) total number density associated with the crossing of the plasma sheet (movement towards the lobe). These values correspond very well with the compiled data of *Kivelson et al.* (2004), who indicated that the average range of density should be between 0.01 to 0.5 cm$^{-3}$, with an AMU of 16 in the current sheet, and AMU of 2 in the lobes. The main difference is that our lowest densities are higher than the lowest densities observed by Galileo. This is because in the simulations the data is taken at a fixed point in the rotational equator whereas the Galileo data has substantial deviations from the equator. This means the spacecraft is actually sampling deeper into the lobes than the one fixed point that we are using for the present discussions. Note also that the overall density is smaller for the low solar wind density case, though the AMU is slightly higher in the tail. This result indicates that even at Calisto's orbit, increased solar wind forcing can lead to a relative enhancement of light ions.

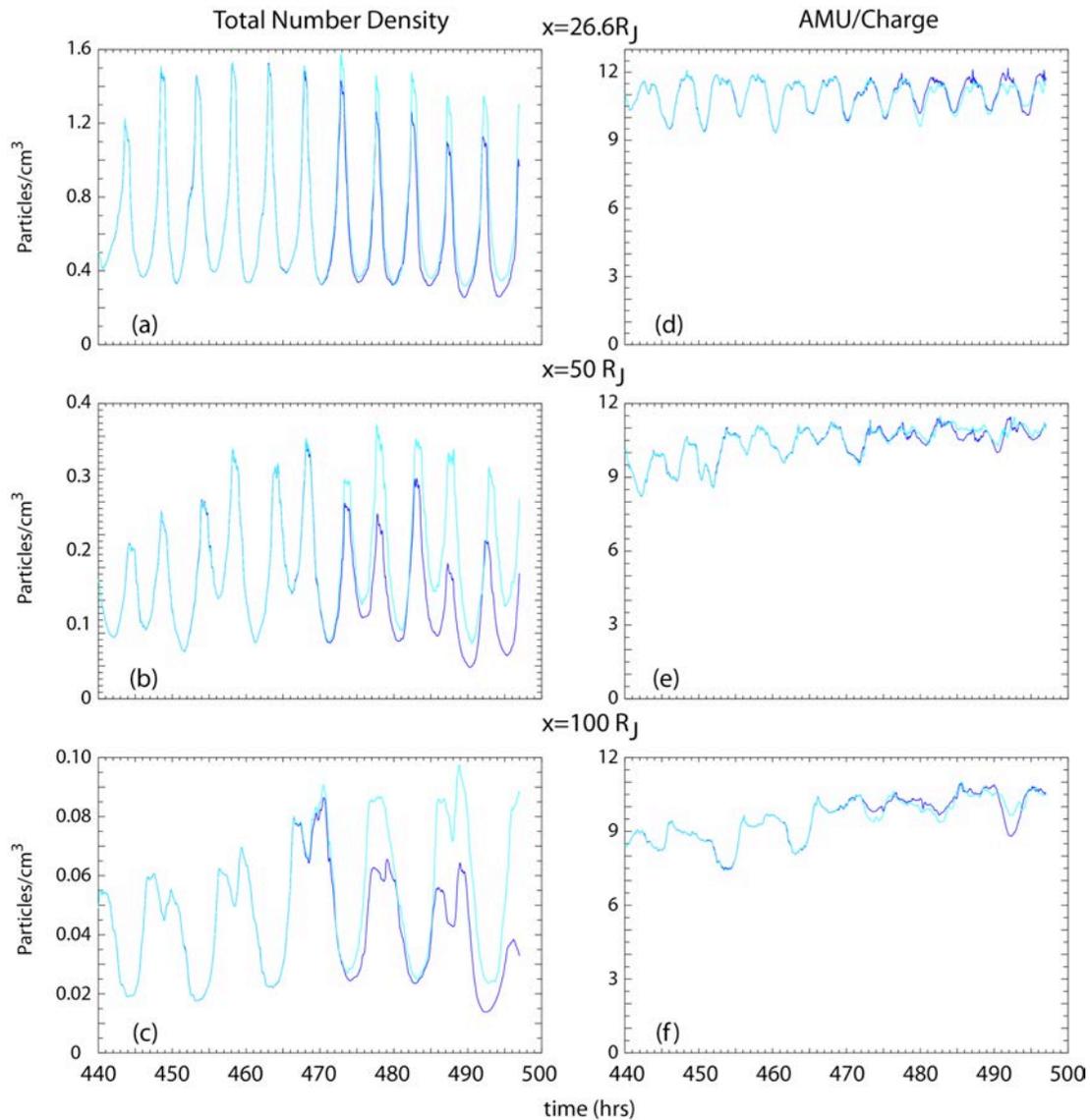

**Figure 6.** The time histories of the variations of the plasma density (left) and average atomic mass unit for three positions in the geographic equator moving down the tail along the noon-midnight meridian. The colors are the same as in Figure 3 with the high solar density case in light blue and the low solar wind density case in dark blue. Note that period of the density fluctuations is seen to double between the inner position and the furthest position down the tail.

Similar to the data for the other positions in the inner magnetosphere, the density time profiles have a period equal to half the planetary period. The value of half comes from the fact that there is a swing in the density as the southern and northern hemispheres wobble through the observing position. As one moves down the tail, the number density and average AMU continue to decline and that the low solar wind density case is associated with the smallest number densities in the outer magnetosphere. More importantly one sees a shift in the periodicity of the density profiles in that at 50 $R_J$ downtail, one of the current sheet encounters tends to be fractionally weaker than the other and by 100 $R_J$ instead of encountering the current sheet twice, there is only one strong encounter so that the density profile appears to have a period that matches the planetary period, as opposed to having twice the frequency as seen in the inner/middle magnetosphere.

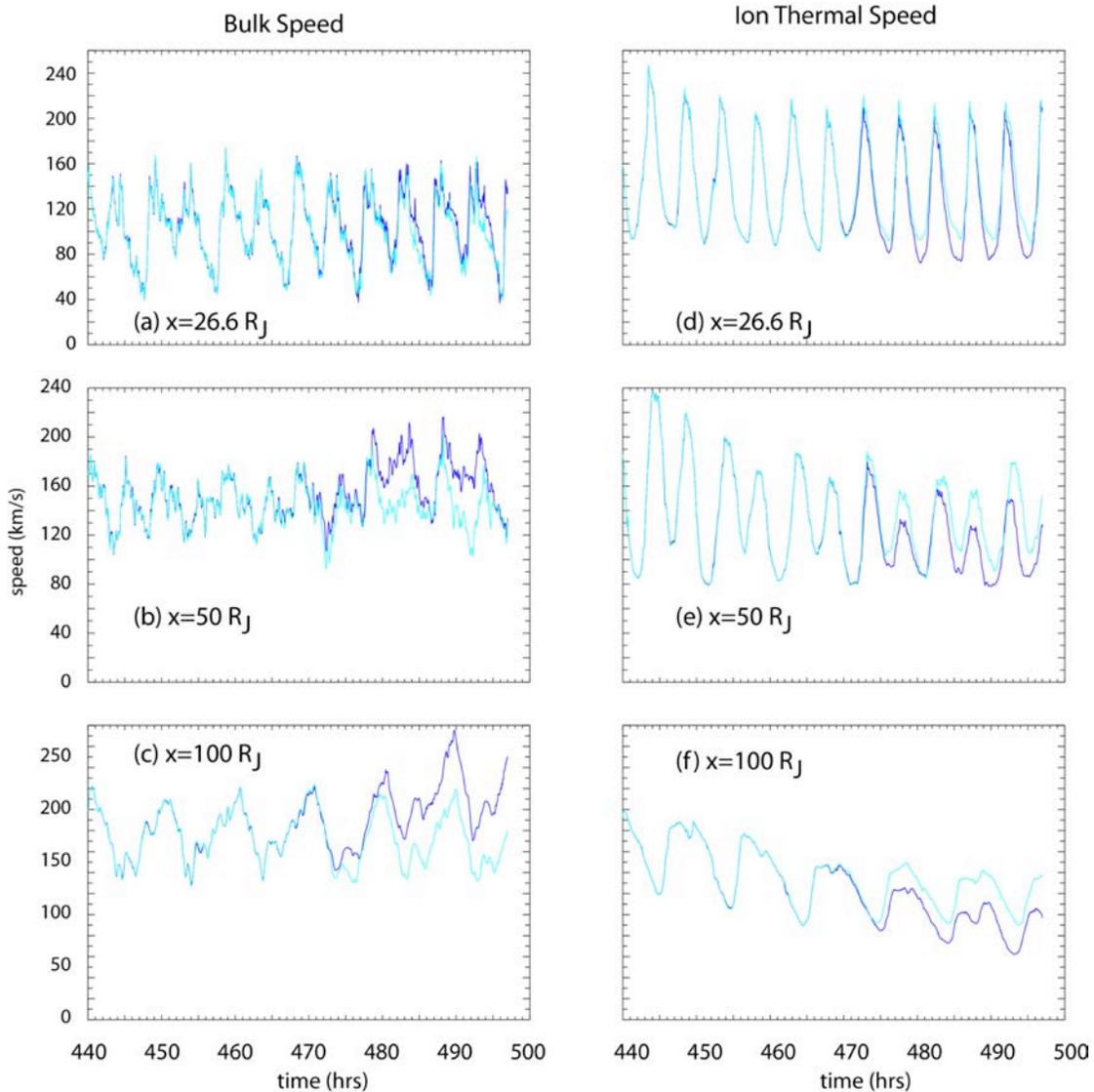

**Figure 7.** The bulk speed (left) and ion thermal speed corresponding to the density profiles in Figure 6. The periodicity is also seen to change with distance down the tail culminating in a period equal to the planet's rotational period. The bulk velocity of the plasma is seen to increase out to 100 $R_J$ with the local solar wind case have the higher bulk speed but lower thermal speed.

The plasma speed (defined as the magnitude of the sum of the momentum vectors divided by the sum of the mass densities) at the same locations shown in Figure 6, are shown on the left hand side of Figure 7. Like that seen in Figure 6, the periodicity is seen to change with distance down the tail culminating in a period equal to the planet's rotation period at the largest distance shown. The speed range that is seen within the simulation model covers the low end of the velocity range noted by *Kivelson et al.* (2004). This lower range may be a local time effect since the addition of a two cell convection pattern with the rotational velocity field is expected to produce higher speeds in the dawn sector and lower speeds in the dusk sector. Similar to the density profiles, oscillations at the second harmonic of the Jovian rotation frequency are seen in the inner magnetosphere, while changes to oscillation that is at the rotation frequency are also present in the outer magnetosphere. An additional effect that is seen in Figure 7 is that the low solar wind density case leads to the observation of higher speeds than the higher solar wind pressure case after about x = 50 $R_J$. This effect is different from the more familiar terrestrial case where strong solar wind forcing tends to produce higher velocities in the distant tail. This effect (analyzed in depth in Section 5) is shown to be localized to the region between 50 $R_J$ and 200 $R_J$. Beyond this region, high speeds for stronger solar wind forcing are seen, similar to the terrestrial magnetosphere.

The ion thermal speed (defined as the square root of the total plasma pressure divided by the total mass density) is shown on the right Figure 7. A transition is seen where the thermal speed goes from exceeding the bulk speed to being below the bulk speed occurs around 50 $R_J$. Irrespective of this transition, the ion thermal speed shows the same periodicity as the bulk speed, though the low solar wind density case is associated with lower thermal speeds beyond the transition point.

As demonstrated in the following, both the change in the apparent periodicity and the differences in the plasma properties between the high and low solar wind density cases are due to the wobble of the Jovian magnetic field, and this wobble response makes the Jovian magnetosphere distinct from the terrestrial and Kronian magnetospheres.

## 5. Outer Magnetospheric Structures: Tail Reconnection

The calculated magnetic field for the three positions is shown in Figure 8. Only the $B_x$ and $B_z$ components are shown for clarity. The range in the magnitude of the magnetic field is consistent with *Kivelson et al.* (2004). At Calisto's orbit, $B_x$ is almost purely sinusoidal in nature with a small perturbation being produced by changes in the solar wind dynamic pressure. $B_z$ remains negative and reaches its average value of -6 nT during the current sheet crossings i.e. when $B_x$ goes to zero. When the magnetic equator is above the observing point (i.e. $B_x < 0$) $B_z$ becomes more negative while for $B_x > 0$, $B_z$ becomes less negative. Thus, wobble of the magnetic field leads to an asymmetric response between the two hemispheres.

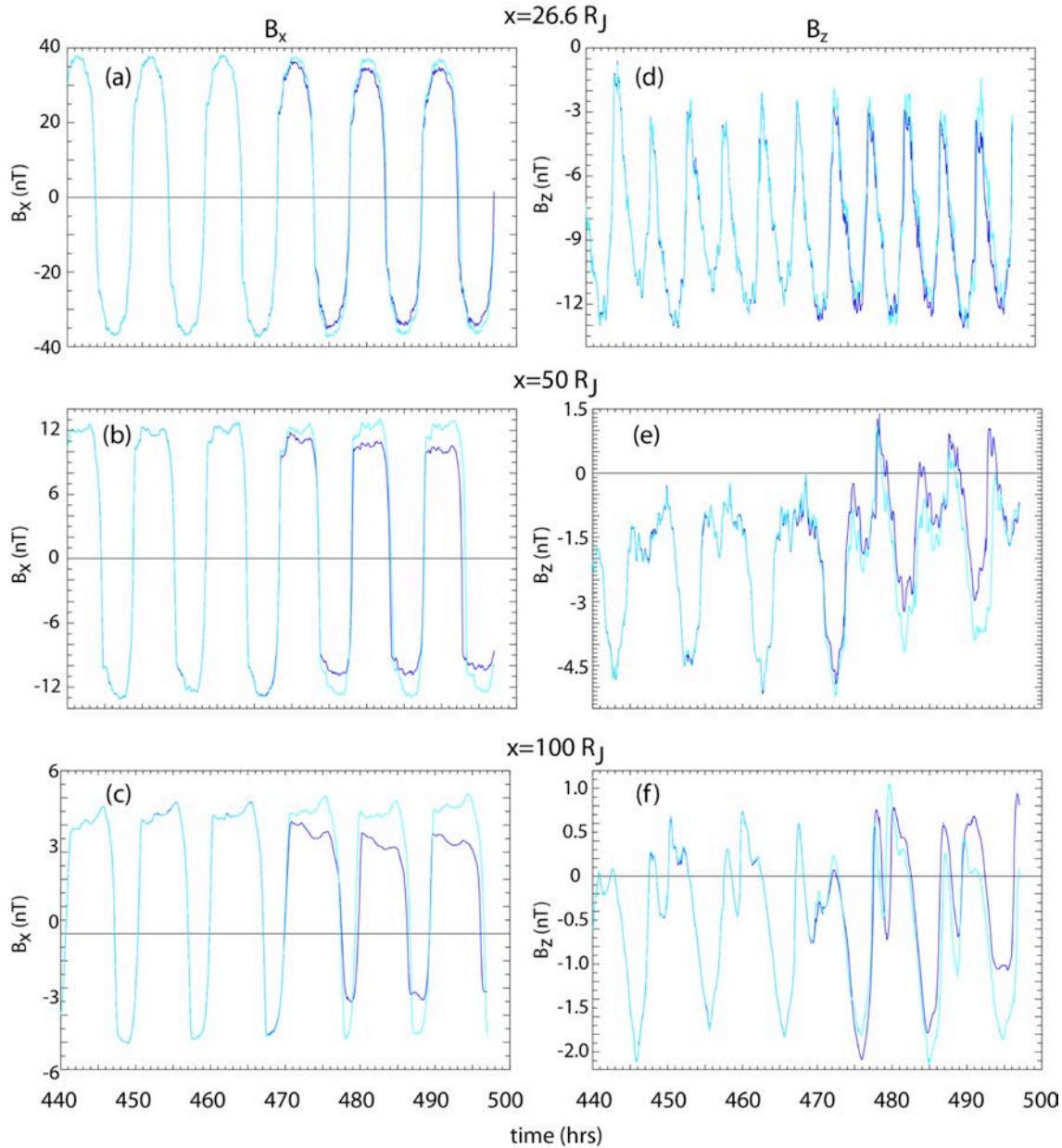

**Figure 8.** The $B_x$ (left) and $B_z$ (right) components of magnetic field for three different positions starting at Calisto's orbit (top) and moving further down tail.

This difference between the hemispheres is an important factor in accounting for the change in periodicity in the plasma density seen in the previous figures. In particular, it is seen in Figure 8b and c that the modulation in the $B_x$ changes from sinusoidal to square wave at x = 50 $R_J$ and to an asymmetric square wave pattern at x = 100 $R_J$. At the same time, positive excursions in $B_z$ develop and are actually strongest for the low solar wind density case. This enhancement occurs because the magnetosphere is larger for low solar wind dynamics pressure and the flapping of the tail current sheet produce by the wobble is corresponding larger.

The presence of positive $B_z$ at Jupiter can be a sign of reconnection and in fact this is the case in the present work. Figure 9 shows a series of snapshots of the magnetic field mapping the outer magnetosphere over a Jovian rotation for the low solar wind pressure case. In the first frame, at t = 474.5 hrs, the magnetosphere has the typical configuration of a fast rotating magnetosphere with closed field lines on the dusk side, and open field lines on the dawn side in the middle and outer magnetospheres. At t = 476.4 hrs, a flux rope starts to develop with a strong core field and has an azimuthal extent of a few hrs in local time. This flux rope then expands in local time to over 6 hrs with a width of near 100 $R_J$. This flux rope is then ejected tailward in the span of a few hours. Once ejected, the magnetosphere returns to the same appearance as in the beginning of the cycle. The appearance of this flux rope corresponds to the time when a positive $B_z$ seen at the individual probe points in Figure 8.

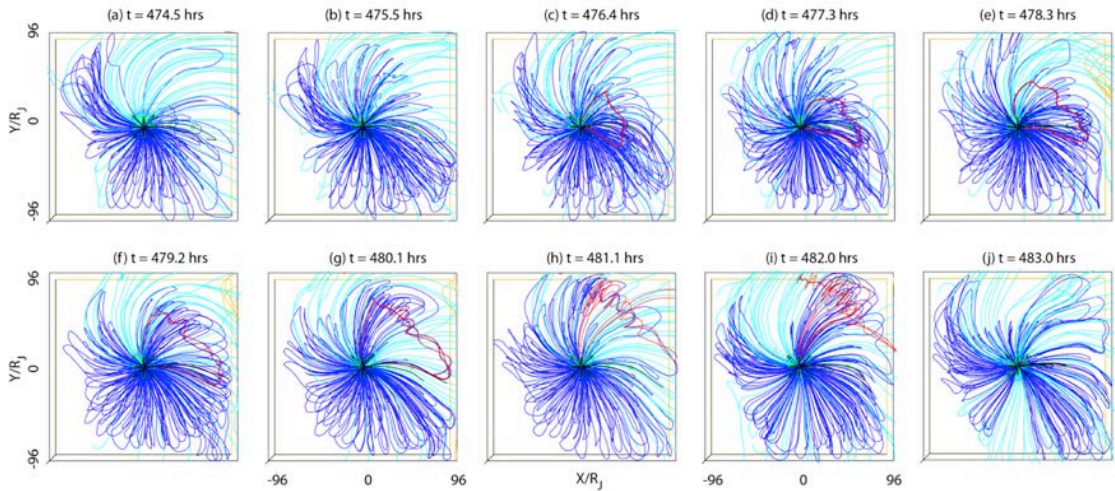

**Figure 9.** Mapping of the Jovian magnetic field lines during the ejection of a flux rope, which is shown in red for the low solar wind pressure case. Dark blue field lines are closed (attached to the northern and southern hemispheres) while the light blue lines are only attached to one of the hemispheres and open at the other end to the solar wind. The orange field lines are solar wind field lines and are not connected to Jupiter. Ejection of the flux rope coincides with the positive excursion in $B_z$ seen in Figure 8.

It is difficult to fully image the spatial extend of the ejected flux rope due to the fact that it is not flat but rather highly curved from the flapping of the tail produced by the wobble of the magnetic field. This effect is illustrated in Figure 10, which shows slices of the ion thermal speed in both the equatorial plane and in the noon-midnight meridian. At the first time shown, the plasma sheet (as identified by the position of the hottest plasma) shows a distinctive wave-like characteristic in the noon-midnight meridian. This wave-like feature has a spatial half-period of about 130 $R_J$ down the tail as seen in subsequent frames, and propagates out of the field of view over a period of a planetary rotation.

The interpretation of the equatorial contours are a little more complicated. Local maxima that appear in cuts through the equatorial plane can be due to either an out-of-plane feature associated with flapping current sheet or they can be discrete features association with a structure such as a flux rope. If the local maximum occurs in both the noon-midnight meridian and in the equatorial plane then they typically associated with distinct structures. If they appear only in the equatorial plane then they are typically associated with the flapping of the current sheet, though this is not always the case if the structure does not fully extend to the noon-midnight meridian.

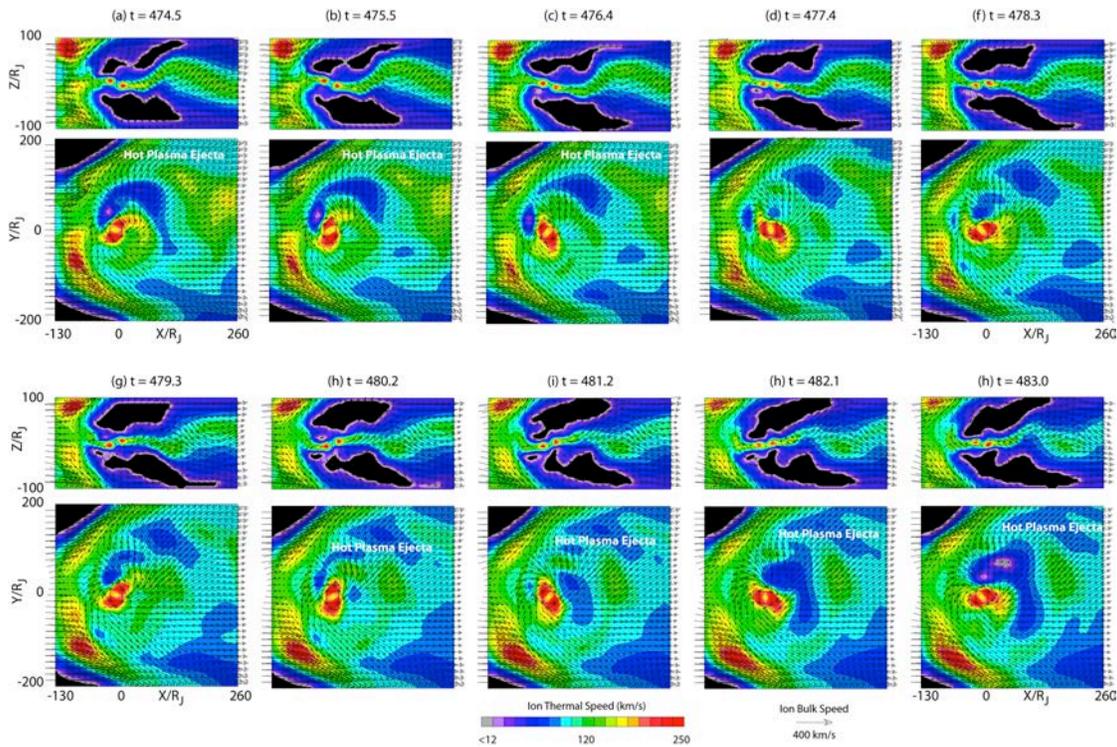

**Figure 10.** Evolution during one Jovian rotation of the ion thermal speed in the noon-midnight meridian (rectangular panels) and in the equatorial plane (square panels). The ejection of the flux rope in Figure 9 is association with the formation and ejection of hot plasma in the magnetotail.

In Figure 10 there is a large region of hot plasma ejecta (high thermal velocity) that is seen in the first few frames at x > 130 $R_J$. This position coincides with the position of the flux rope in Figure 9. This feature is seen to expand in azimuth and propagate down the tail. This hot plasma ejecta associated with a local maximum in both the equatorial plane and in the noon-midnight meridian. Due to the rotation of Jupiter it propagates both down the tail as well as towards the dawnside. The thermal velocity in the equatorial plane regions a maximum around t = 481 hrs near 200 $R_J$, indicating that the flux rope ejection is also associated with local heating of the plasma.

Another way to look at the ejection of the flux rope is through the contours of the Alfven speed relative to the bulk speed (i.e. the Alfvén Mach number) that are shown in Figure 11. The solar wind and magnetosheath appear as white, being super-Alfvénic. The magnetopause is seen as the region between these super-Alfvénic regions and the interior sub-Alfvénic region. The magnetopause is neither smooth nor static due to wobble of the plasma current sheet. The center of the current sheet, where the magnetic field tends to be weakest within the magnetosphere, is easily identified by high values of Alfvén Mach number within the magnetosphere. The interception of the plasma sheet with the equatorial plane creates a spiral pattern in the inner and middle magnetosphere associated with the reduction in the rotation rate of the plasma at large distances. Note that in the inner/middle magnetosphere, the spiral has two arms associated with cuts through the current sheet on opposing local times. The axis of the inner two parts of the spirals is orthogonal to the dipole axis. In the noon-midnight meridian the current sheet has substantial curvature from the wobbling of the Jovian magnetic field.

At the first time shown (t = 474.5 hrs) the current sheet arm that attaches to the dawnside of the inner magnetosphere has an extension all the way across to the duskside in the middle magnetosphere. The duskside arm that attaches to the inner magnetosphere extends back along the dayside magnetopause at this stage, and has the lowest Alfvénic Mach number of the two arms. At the next time shown (t = 476.4 hrs) the inner portions of the two arms have increasing Alfvénic Mach number and, more importantly, the leading edge of the dusk arm is gaining on the trailing portion of the dawn arm. These two sections then start interacting at t = 478.2 hrs along a large front in the night side extending from the dusk to past midnight. It is at this time that the flux rope of Figure 9 starts to strongly develop and to be ejected down the tail. This difference in rotation rate between the two hemispheres of the magnetosphere is the driver of the change in the periodicity from the inner to outer magnetosphere and, as will be discussed in Section 8, is a quintessential feature of a fast rotating magnetosphere with a tilted dipole.

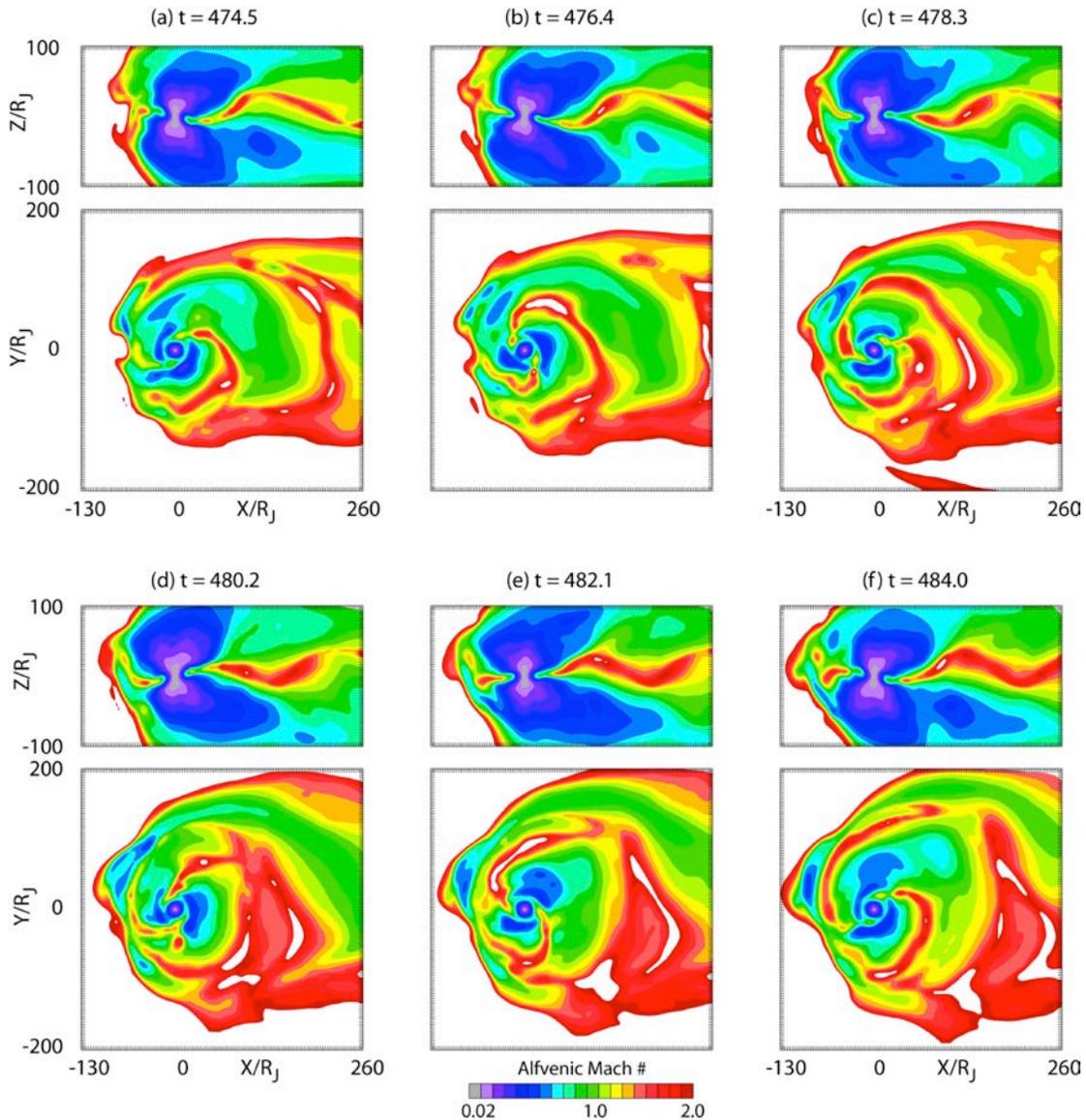

**Figure 11.** Contours of the Alfvénic Mach number covering the same period as in Figure 10, except the cadence is at 1.9 hrs instead of 0.95 hrs. The center of the current sheet is characterized by super-Alfvénic flows.

## 6. Distant Tail Features.

The propagation of some of the features associated with a flux rope into the distant tail is illustrated in Figure 12, which shows contours of the current intensity in the noon-midnight meridian for the high and low solar wind density cases for one planetary period. The full width at half height of the tail is about 2-4 $R_J$ at 200-400 $R_J$ down the tail. The density in the current sheet in this region is < 0.1 cm$^{-3}$ which yields an ion skin depth of > 10$^4$ km. At this value non-ideal MHD effect can be important in determining reconnection rates.

Due to the wobble, the current sheet has a distinct flapping motion down the distant tail, with the flapping still present beyond 600 $R_J$, although the magnitude of the flapping does decrease with distance down the tail. The red-dashed line shows the position of the trough moving out for the high density case. The corresponding phase speed is about 600 km/s. The white-dashed line indicates the movement of the trough for the low density case. Its velocity is about half as much as the high solar wind density case. However, the magnitude of the flapping is larger for the low solar wind density case by as much as 20-30 $R_J$. Thus for a spacecraft moving down the tail, the probability of seeing an enhanced energy flux is high for low solar wind conditions, though the energy of the particles that would be detected would be lower. This dependence on the solar wind conditions could explain the deep tail observations of energetic particle events observed by New Horizons (*McComas et al.*, 2007; *McNutt et al.*, 2007) which have both a 10-hr periodicity as well as longer duration periodicity of 3 days. Here the 10-hr periodicity is driven by the wobble while the 3-day period would be driven by solar wind conditions.

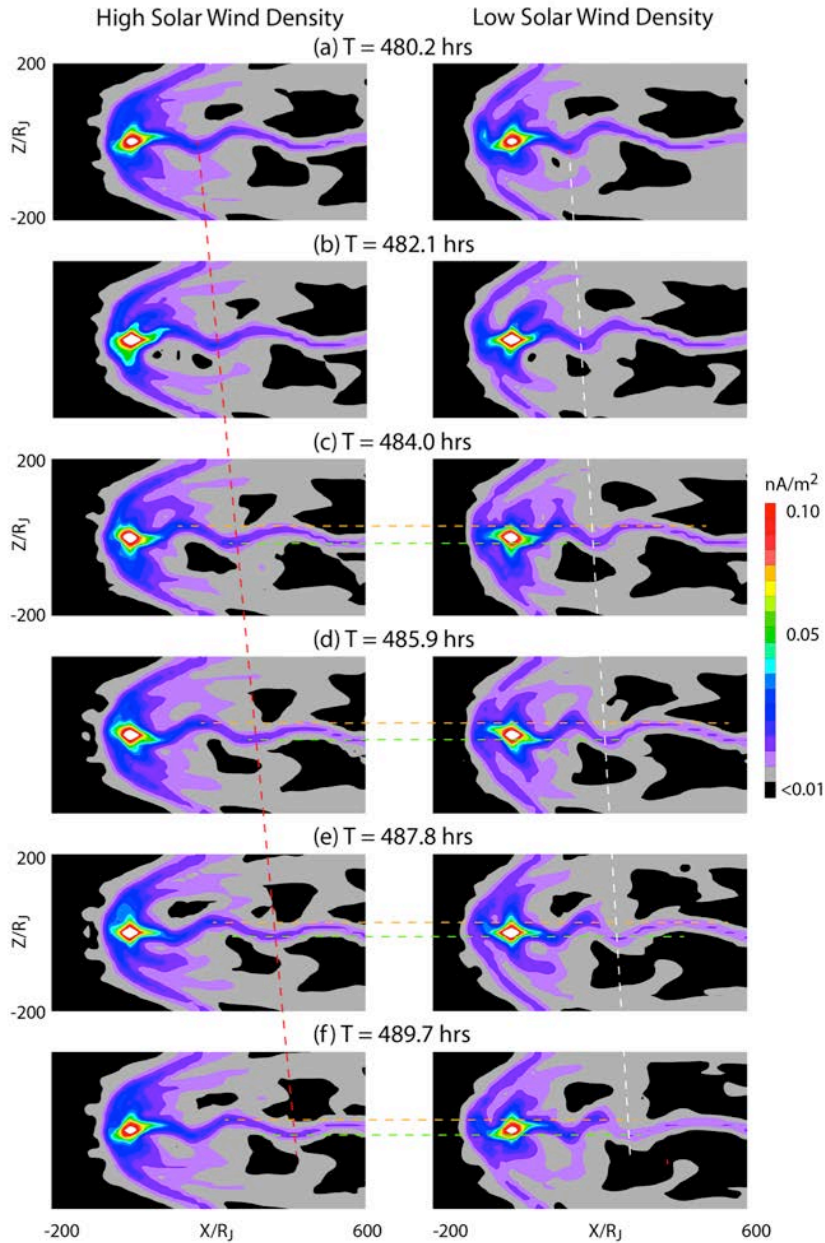

**Figure 12.** Current density along the noon-midnight meridian for the high (low) solar wind case on the left (right) hand side. The vertical red dashed lines indicate the position of the first trough in the tail current sheet assuming it is moving out a constant velocity. The horizontal dashed lines mark the center of the current sheet at the first peak and trough for the high density solar wind case. The flapping of the current sheet for low-density solar wind case is larger for the low-density solar wind case but its average speed in the distant tail is lower.

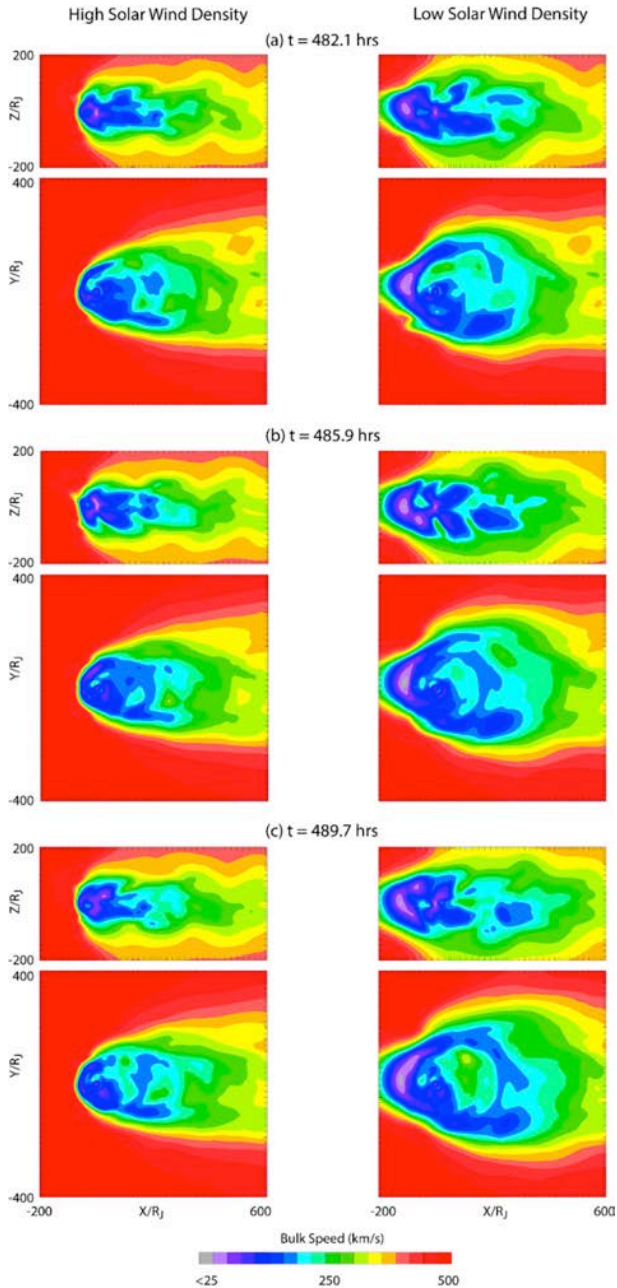

**Figure 13.** Left (Right) contours of the plasma bulk velocity in the noon midnight meridian and in the equatorial plane for the high (low) solar wind density case.

While the phase speed of the flapping is higher than the solar wind speed, the group speed or plasma bulk speed is much lower as shown in Figure 13, which plots contours of the plasma bulk speed for the two different solar wind conditions in both the noon-night meridian and equatorial planes. In both cases the bulk speed of the plasma in the tail does not reach solar wind speeds until distances greater than 600 $R_J$. While there are local regions where the speed in the outer magnetosphere, for the low-density solar wind case, exceed those seen in the high density case, the plasma speed in the distant tail is on average slower in the low density case than the high density solar wind case. Result indicates that solar wind forcing is still an important factor in the energization of the plasma in the distant tail. And while the high solar wind case has the higher speed on average, flux rope ejection can lead to local regions of high speed plasma for the low-density case that exceeds that seen in the high density case.

Furthermore it is seen that each of the ripples in the tail current sheet leads to ripples of high-speed plasma along the plasma mantle of the magnetosphere. Flapping of the current sheet also leads to intrusion of high-speed flows at high latitudes. The amplitude of these penetrations appears largest for the low-density solar wind case.

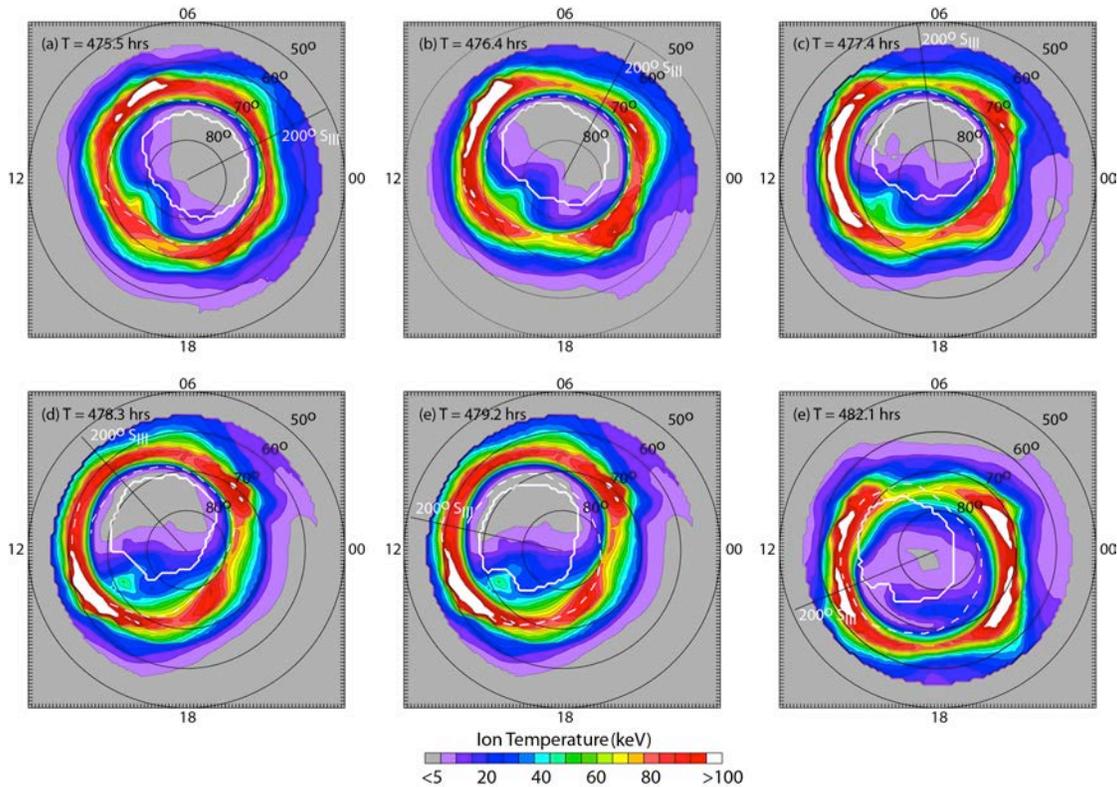

**Figure 14.** The mapping of the hot ion population in the magnetosphere onto the auroral regions. The latitudes are relative to the rotation axis with the sun at noon local time. The position of the dipole is given by the 200° latitude line. The solid white line solid indicates the open/closed boundary as identified by field line mapping. The dashed line shows the position of the peak in the plasma azimuthal velocity relative to the rotation velocity, which is an approximate measure of the position for the breakdown of plasma co-rotation.

## 7. Auroral Processes.

The mapping of the processes that are tied to the auroral region is illustrated in Figure 14 for the high density solar wind case. In this figure, field lines from the auroral zones are mapped into the magnetosphere and the hottest ion populations along the field line are then plotted as a function of local time and latitude. If one also assumes that the precipitating particles originate from hot magnetospheric populations then this mapping provides a pseudo-representation of possible auroral features. Also shown are the positions of the open/closed boundary as defined by whether the field maps map to the Jovian ionosphere or out into the solar wind (solid white line), and the position of the breakdown of the co-rotation as defined by the peak in the azimuthal velocity relative to the co-rotation velocity (dashed white line). The position of the dipole axis is given by the 200° longitude line. Much like the terrestrial magnetosphere, there are three main ion acceleration processes incorporated in the modeling: (a) centrifugal acceleration of plasma moving outwards from low altitudes to large radial distances (augmented by the fast rotation of Jupiter, (b) current sheet acceleration (including reconnection) and (c) betratron acceleration as plasma is convected from the tail inwards into stronger magnetic field.

At the first time shown in Figure 14, the hot populations in the distant tail associated with the ejection of the flux rope at times t = 474.5-476.4 from Figures 9 and 10 do not appear in the auroral mapping as the field lines that they populate have already become detached from the auroral ionosphere. Instead the hottest populations are associated with the inner magnetosphere at slightly lower latitudes than the co-rotation boundary, and this region corresponds to the stable main auroral oval (e.g., *Grodent et al*., 2003). The coldest ions appear at higher latitudes aligned with the magnetic axis. This cold region corresponds well with the dark regions of the dawnside auroral oval (e.g., *Grodent et al*, 2003). This region has the lowest energy particles tied to it when the dipole axis is pointed to the dawnside, with some intensification at high latitudes and a weakening of the main auroral oval when the dipole is pointed to the dusk sector.

As a precursor to the ejection of the flux rope, a hot population is seen to moving from the pre-noon sector across to the post-noon sector (Figures 14a-c). With the arrival of this hot population at about 15LT (Figure 14d) an intensification of a high latitude feature at about 15 LT is seen to develop and move into the high altitude region. This feature is seen to break away from the main oval and cross the open-close boundary. It then propagates towards the night sector over a period of few hours, and fades in intensity. This feature coincides with the ejection of the flux rope seen in Figure 9 and has features similar to the arc-like structures reported of *Grodent et al.* (2003) in terms of local time positions, movement and duration. Here, we note that the fast rotation and wobble of Jupiter makes these storms appear to start on the dusk sector and move towards midnight which is opposite to the situation in the terrestrial magnetosphere.

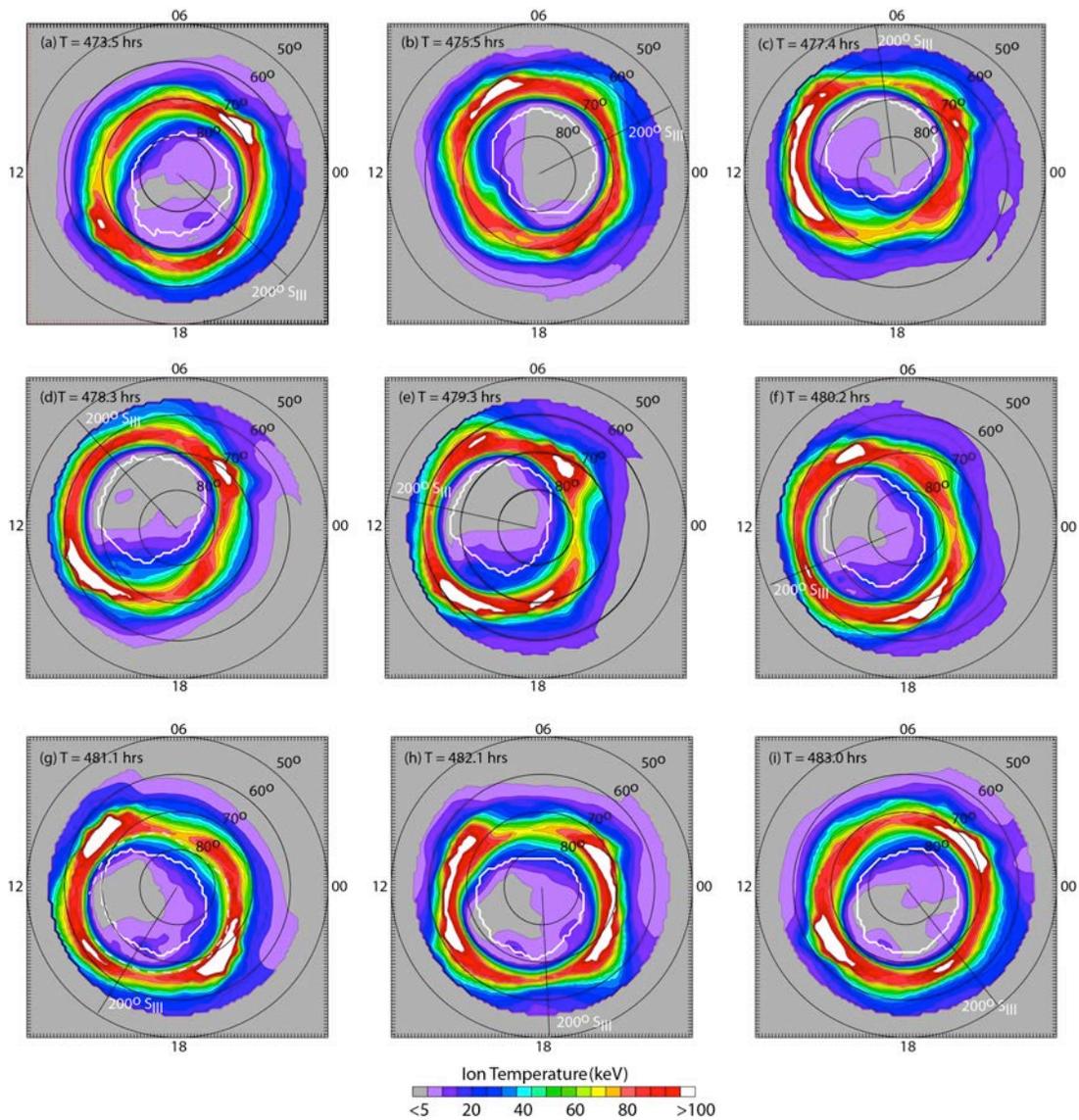

**Figure 15.** As in Figure 14, except for the low density solar wind case, and the time period shown covers a full planetary rotation period.

The corresponding results for the low solar wind density case are shown in Figure 15. The main oval feature has essentially the same position and intensity as the high density solar wind case. This is consistent with the dynamics inside the co-rotation boundary being control by the planets rotation and not by the prevailing solar wind conditions. The formation of the high latitude feature has the same temporal evolution as the high density case but the intensity of the high latitude feature is greatly reduced, consistent with the lower energization of particles in the outer magnetosphere described in the previous sections.

## 8. Summary.

In this paper, we have examined a global Jupiter magnetosphereic model that incorporates several of the different ion species and sources. The model is limited in that the inner radius has to be set to 2 $R_J$ and as a result there are enhanced losses to the inner boundary, which has to be compensated with an increased injection rate of plasma into the Io plasma torus. This is still significantly closer than previous simulations. Also the model is the first to be able to reproduce many of the observed features of the inner magnetosphere to within a factor of two. Much of the density variations at a fixed observing point originate from the wobbling of the torus passed the observing point. Moving out into the outer magnetosphere, the plasma and magnetic fields properties change from being purely sinusoidal to an asymmetric square wave modulation has a period equal to the planet's rotational period. This change originates from the fact that the two hemispheres of the Jovian magnetosphere under fast rotation do not behave identically. As a result, tail reconnection occurs with a periodicity equal to the planetary rotation. This reconnection lead to energetic particles being ejected both down the tail and into the dawnside. In the latter case, these particles are convected around the dayside and into the dusk sector where they participate in further reconnection events. It is shown that the solar wind dynamics pressure does not modify the reconnection period, but does control the energization of the particles involved in the reconnection processes.

The mapping of particles populations derived within the simulations have many features that overlap with observations of the Jovian auroral oval. In particular, the mapping of the particles within the co-rotation boundary corresponds with the position of the main oval, and, like the main oval, the particle populations are stable and not sensitive to changes in the solar wind conditions. Due to the reconnection events, the dawnside populations of outer magnetosphere overlap with the dark regions of the auroral oval. The energetic particles rotating around from noon into the dusk sector and beyond provide a precursor to the tail reconnection and their mapping coincides with the arc-like structures seen in the high latitude auroral regions in the main oval.

A simplified model explaining the importance of the wobble of the Jovian magnetosphere is shown in Figure 16. For the case where the dipole is assumed to be aligned with the rotation axis (Figure 16a) and the solar wind flow is perpendicular to the rotation axis, there is no asymmetry present between the northern and southern hemispheres. However, when the dipole is tilted relative to the rotation axis, then, depending on the position of the dipole axis, the dayside magnetosphere can experience an upward force, if the dipole is pointed into the tailward, and a downward force, if the dipole is pointed into the dayside. At the same time, if there is significant ionospheric outflow, then the opposite forcing occurs on the tail.

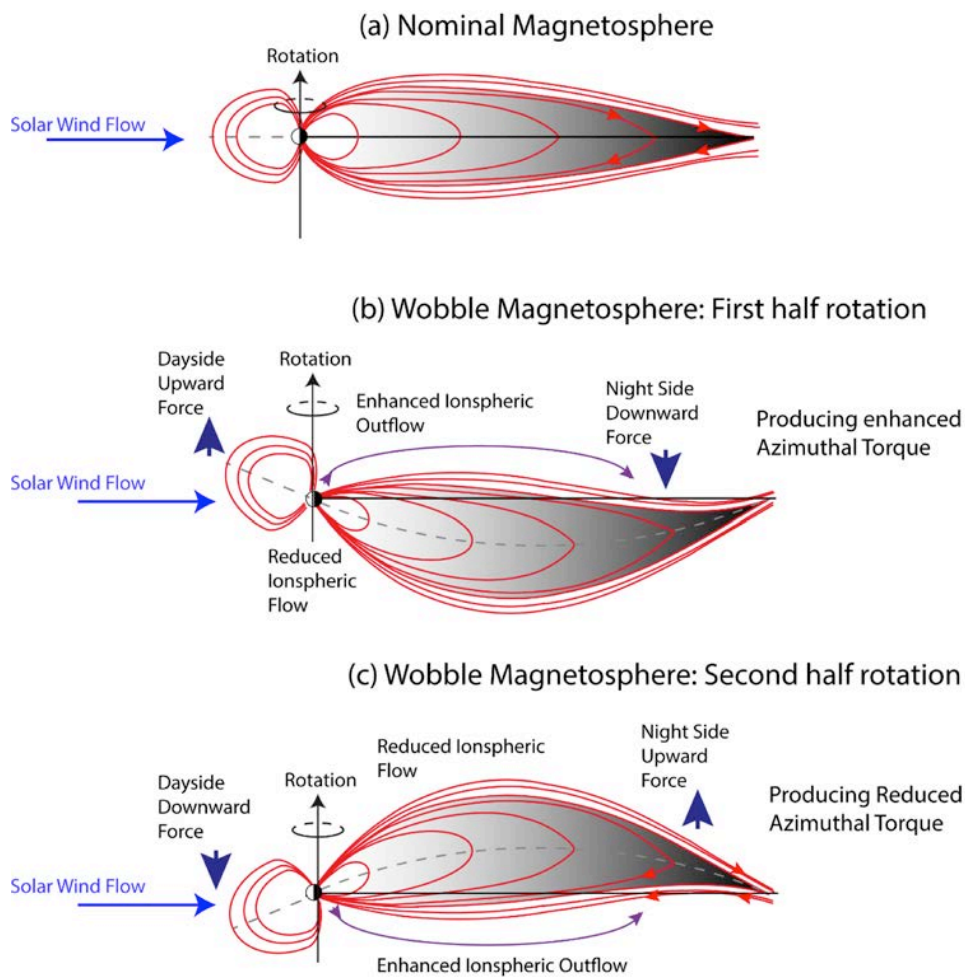

**Figure 16.** Simplified model for the importance of magnetic wobble in fast rotating magnetosphere.

The resultant effects on the magnetosphere then depends on whether the magnetosphere is a fast or slow rotator, and the tilt of the dipole relative to the rotation axis. Whether the planet is a fast or slow rotator is defined by the rotation period relative to the inductive or response period of the magnetosphere. For example the terrestrial magnetosphere is a slow rotator (with a rotation period of 24 hrs and substorm growth phase of only 30-60 min). In this case reconnection occurs below or above the rotation equator.

For fast rotating system, where the plasma transport in the middle/outer magnetosphere is much longer than the rotation period, the forces produced by the solar wind and ionospheric outflows leads to a torque on the system. This torque increases (decreases) the azimuthal velocity for the configuration in Figure 16b (16c). This change in azimuthal velocity is the reason that the current sheets associated with different parts of the magnetosphere are seen to coalesce, leading to local enhancement of the tail current intensity and the development of reconnection. With a tilt angle of 10 degrees about 17% of the incident solar wind force is directed orthogonal to the current sheet. With an angular velocity inside the magnetopause of about half the solar wind speed, the power from the torque is about 10% of the incident solar wind power. As such, the planetary wobble is a significant factor on the dynamics of the magnetosphere.

It should be noted that this model is simplified in that the large size of the Jovian magnetosphere (particularly during low solar wind dynamics pressure) means that both the dayside and nightside can have multiple ripples in its current sheets as opposed to the single ripple shown in Figure 16. Nevertheless, presence of an overall torque that changes single during a rotation will still be present.

In summary, the Jovian system differs from the terrestrial system in that torques, arising from the interaction of the solar wind and ionospheric outflows with the rotation of a tilted dipole, lead to differential motion between its northern and southern hemisphere. This differential motion then drives reconnection at the planetary rotation period. It also differs from the terrestrial magnetosphere in the mapping of the position of energetic particles at high altitudes is towards the dusk sector as opposed to midnight in the terrestrial magnetosphere.  Solar wind conditions are shown to drive the intensity of the particular acceleration in the outer magnetosphere, but the reconnection rate is shown not be insensitive to this very important factor for the terrestrial magnetosphere.

**Acknowledgements.** The authors acknowledge helpful discussions supported by the International Space Science Institute Workshop on Coordinated Numerical Modeling of the Global Jovian and Saturnian Systems.